# Financial Risk-Based Scheduling of Microgrids Accompanied by Surveying the Influence of the Demand Response Program

Tohid Khalili, *Student Member, IEEE*, Hamed Ganjeh Ganjehlou, Ali Bidram, *Senior Member, IEEE,* Sayyad Nojavan, Somayeh Asadi

***Abstract*—This paper presents an optimization approach based on the mixed-integer programming (MIP) to maximize the profit of the Microgrid (MG) while minimizing the risk in profit (RIP) in the presence of demand response program (DRP). RIP is defined as the risk of gaining less profit from the desired profit values. The uncertainties associated with the RESs and loads are modeled using normal, Beta, and Weibull distribution functions. The simulation studies are performed in GAMS and MATLAB for 5 random days in a year. The simulation results show that RIP is reduced when downside risk constraint (DRC) is considered and DRP is implemented. Although DRP increases the total profit of the MG, it also notably increases the risk. On the other hand, considering DRC significantly reduces the percentage of the risk with a slight decrease in the profit.**

***Index Terms*—Demand response program, microgrid, optimization, profit, risk assessment.**

## NOMENCLATURE

### Indices and Sets

| | |
|---|---|
| *bat* | Battery |
| *ch* | Charge |
| *disch* | Discharge |
| *g* | Index of DGRs |
| *i* | Counter of PVs number |
| *j* | Counter of WTs number |
| *p* | Profit |
| *r* | Risk |
| *RIP* | Risk in profit |
| *s* | Index of scenario |
| *t* | Index of hour |
| *w* | Counter of buses number |

### Parameters and Constants

| | |
|---|---|
| $\alpha$ , $\beta$ | Beta's parameters |
| $\eta^{ch}$ | Charging efficiency |
| $\eta^{disch}$ | Discharging efficiency |
| $\mu$ | Average value |
| $\sigma^2$ | Variance value |
| $a$ | Lower limit |
| $b$ | Upper limit |
| $b_g$ | Coefficient of DGRs' cost function [$/kW] |
| $c_g$ | Coefficient of DGRs' cost function [$] |
| $c_1$ , $k_1$ | Weibull's parameters |
| $down\ rate_g$ | Minimum rate of decrease in the DGRs power [kW] |
| $M_p$ | A large and positive number |
| $P_g^{\max}$ | Maximum generation capacity of the $g^{th}$ DGR [kW] |
| $P_g^{\min}$ | Minimum generation capacity of the $g^{th}$ DGR [kW] |
| $P_{\min}^{ch,bat}$ | Minimum charging power of the BESS [kW] |
| $P_{\max}^{disch,bat}$ | Maximum discharging power of the BESS [kW] |
| $PV_{i,t,s}^{\max}$ | Maximum generation capacity of the $i^{th}$ PV [kW] |
| $q$ | A real number |
| $shut_g$ | Shutdown cost of DGRs [$] |
| $SOC_{\min}$ | Minimum permissible value of the BESS SOC |
| $SOC_{\max}$ | Maximum permissible value of the BESS SOC |
| $start\ up_g$ | Startup cost of DGRs [$] |
| $up\ rate_g$ | Maximum rate of increase in the DGRs power [kW] |
| $WT_{j,t,s}^{\max}$ | Maximum generation capacity of the $j^{th}$ WT [kW] |

### Functions and Variables

| | |
|---|---|
| $\lambda_p$ | A number between 0 and 1 |
| $\rho(t)$, $\rho(k)$ | Power price at $t^{th}$ and $k^{th}$ hours [$] |
| $\rho_0(t)$, $\rho_0(k)$ | Initial power price at $t^{th}$ and $k^{th}$ hours [$] |
| $A(t)$, $A(k)$ | Customer incentive at $t^{th}$ and $k^{th}$ hours [$] |
| $b_{t,s}^{bat}$ | Binary variable |
| $C_{t,s}^{sell}$ | Selling price of the electricity [$] |
| $C_{t,s}^{buy}$ | Purchase price of the electricity [$] |
| $Cost_{DRP}$ | Total cost paid as an incentive [$] |
| $\overline{EDR}_p$ | Expected downside risk (EDR) of the system |
| $E(t,t)$ | Self-elasticity at $t^{th}$ hour |
| $E(t,k)$ | Cross elasticity between $t^{th}$ and $k^{th}$ hours |
| $f(v)$ | Weibull distribution function |
| $g(y')$ | Beta distribution function |
| $h(q,\mu,\sigma^2)$ | Normal distribution |
| $P_{g,t,s}$ | $g^{th}$ DGR's generated power at $t^{th}$ hour and $s^{th}$ scenario [kW] |
| $P_{trans,t,s}$ | Transmission line power [kW] |
| $P_{t,s}^{ch,bat}$ | BESS charged power [kW] |
| $P_{t,s}^{disch,bat}$ | BESS discharged power [kW] |
| $P_{t,s}^{buy}$ | Purchased power from the main grid [kW] |
| $P_{t,s}^{sell}$ | Sold power to the main grid [kW] |
| $Pb_{w,t,s}$ | Injected power to $w^{th}$ bus [kW] |
| $pen(t)$, $pen(k)$ | Consumer penalties at $t^{th}$ and $k^{th}$ hours [$] |
| $PL_{p,t,s}$ | Demanded load [kW] |
| $PL_{p,t,s}^{D}$ | Reduced/increased load by the consumer [kW] |
| $PL_{p,t,s}^{LD}$ | Load after the implementation of the DRP [kW] |
| $PL'^{D}_{p,t,s}$ | Reduced load by the customer [kW] |
| $prob_s$ | Possibility of each scenario |
| $prob_{p,s}$ | Probability of the $s^{th}$ scenario |
| $profit_s$ | Profit at the $s^{th}$ scenario [$] |
| $PV_{i,t,s}$ | Generated power of the $i^{th}$ PV [kW] |
| $risk_{p,s}$ | Risk value at the $s^{th}$ scenario |

Tohid Khalili and Ali Bidram are supported by the National Science Foundation EPSCoR Program under Award #OIA-1757207.
Tohid Khalili and Ali Bidram are with the Department of Electrical and Computer Engineering, University of New Mexico, Albuquerque, USA. (e-mails: {khalili, bidram}@unm.edu). Hamed Ganjeh Ganjehlou is with the Faculty of Electrical and Computer Engineering, University of Tabriz, Tabriz, Iran. e-mail: (hamed_mganjehlo@aut.ac.ir). Sayyad Nojavan is with the Department of Electrical Engineering, University of Bonab, Bonab, Iran. e-mail: (sayyad.nojavan@ubonab.ac.ir). Somayeh Asadi is with the Department of Architectural Engineering, Pennsylvania State University, University Park, USA. e-mail: (sxa51@psu.edu).

| | |
|---|---|
| $shut_{g,t}^{cost}$ | Shutdown cost of $g^{th}$ DGR [$] |
| $SOC_{t,s}$ | SOC of the BESS at $t^{th}$ hour and $s^{th}$ scenario |
| $SS_{g,t,s}$ | Binary index |
| $start\ up_{g,t}^{cost}$ | Startup cost of $g^{th}$ DGR [$] |
| $target_{p,s}$ | Target value at the $s^{th}$ scenario |
| $v'_{g,t,s}$ | Binary index of the UC status |
| $W_{p,s}$ | Binary index |
| $w_{r,p}$ | Profit value without considering downside risk [$] |
| $WT_{j,t,s}$ | Generated power of the $j^{th}$ WT |
| $X_{t,s}^{grid}$ | Binary variable |
| $y_{g,t,s}$ | Binary index |
| $\overline{Z}_{profit}$ | Average profit of the MG [$] |
| $\overline{Z}_{profit}^{DRP}$ | Average profit of the MG by considering the DRP [$] |
| $ZZ_s^{profit}$ | Profit of each scenario [$] |
| $ZZ_s^{profit,DRP}$ | Profit of each scenario in this specific case [$] |

## I. INTRODUCTION

MICROGRIDS (MGs) have gained much attention due to their improved reliability and resilience. Moreover, they facilitate the integration of renewable energy sources (RESs) and energy storage systems (ESS) [1]. The optimized operation and energy management of microgrids are of particular importance which have been widely investigated in the literature. The control and operation of MGs are investigated in [2]. In [3], an MG energy management scheme is proposed. In some studies, a single objective is considered to accommodate the optimized operation of MGs. For example, in [4], the performance of the energy management of an MG is optimized. Scheduling of an MG integrating battery energy storage systems (BESSs), fuel cells, wind turbine (WTs), photovoltaics (PVs), and micro-turbines (MTs) is investigated in [5]. Moreover, the reliability of the power system is analyzed in [6]. In [7], power scheduling considering economic and environmental aspects is performed in an MG with the goal of minimizing MG's total operational costs. Also, a hierarchical framework for the optimal operation of MGs is presented in [8]. In [9], a scheduling model for MG's generation is suggested in which the cuckoo-search optimization algorithm is utilized with the aim of minimizing operating costs with or without the demand response program (DRP). In [10], a novel method is presented for stochastic optimal power flow with DRP and considering the quality of service. The multi-objective performance optimization of MGs is investigated in [11]–[12]. The impact of the incentive-based DRP on the MG's operation is analyzed in [11]. And, [12] shows that DRP is an efficient way to increase the quality of service delivered to customers.

Despite the several advantages of utilizing RESs in MGs, the price of the power purchased from utility is usually lower than the RESs' power price. Additionally, the intermittent nature of RESs imposes more risk in profit (RIP) from the MG operator point of view. RIP is defined as the risk of gaining less profit from the desired profit values. The risks associated with the different uncertain parameters of the power system are studied in [13]. Due to uncertainties associated with RESs, risk in the scheduling and price uncertainties in the power trade is considered. In another approach, the influence of uncertainty of responsive loads on the risk-based optimal operation of a smart MG is analyzed in [14]. Furthermore, a stochastic risk-aware optimization model considering profit and risk is presented in [15]. In [15], to perform hydro, wind, and thermal scheduling, predator and prey strategy evolutionary algorithm is used.

In this paper, MG's optimal power scheduling of RESs, BESSs, and diesel generators (DGRs) is implemented with the goal of RIP minimization and maximizing the MG profit by optimizing the purchased and sold power between the MG and the main grid. Moreover, DRP is utilized to maximize the MG profit. The uncertainties associated with the RESs and loads are modeled using normal, Beta, and Weibull distribution functions. With the proposed approach, the MG operator can maximize its profit by performing optimal scheduling of different resources as well as optimizing the traded power. The optimization is performed using mixed-integer programming (MIP). In the MIP problem, DGRs are dispatched using unit commitment (UC). The simulations are done for five random days of a year using GAMS and MATLAB. The simulation results show that RIP is reduced when downside risk constraint (DRC) is considered and DRP is implemented. Although DRP increases the obtained profit of the MG, it also notably increases the risk. On the other hand, DRC significantly reduces the percentage of the risk with a slight decrease in profit.

The novelty and contributions of this research are briefly described as follows:

- Risk-based scheduling of a grid-connected MG with high penetration of RESs and BESSs is proposed.
- The probabilistic nature of the load, PVs, and WTs are considered for assessing the risk of the MG's operation. Moreover, the power price uncertainty is considered.
- The effect of the DRP on the risk of the trade between the MG and the main grid is taken into consideration for helping the MG's owner to minimize its risk or maximize its profit. Moreover, a sensitivity analysis is performed to examine the impact of the participation rate of DRP on the RIP.

The structure of this paper is organized as follows: Section II models the considered MG, MG's components, and some other parameters. Section III elaborates on the constraints and objective functions. The proposed optimization method has been presented in Section IV. In Section V, the simulation results have been analyzed and presented. Finally, the main conclusions are summarized in Section VI.

## II. MODELING

In this section, the structure of MG is introduced. Then, the models of the input parameters are expressed. This section contains several subsections which will be described briefly.

### A. Structure of MG

The considered MG consists of 6 buses and it has been operated under the grid-connected mode. Four of the PVs are installed on the first bus and two other PVs are installed on the second bus. WTs are located on the third bus and the DGRs are installed in the fourth, fifth, and sixth bus. Also, there are BESSs in the mentioned MG which support the MG in the case of power shortage or overage. BESS and load are installed in the sixth bus. Furthermore, MG's load can participate in DRP. The mentioned MG's schematic is illustrated in Fig. 1.

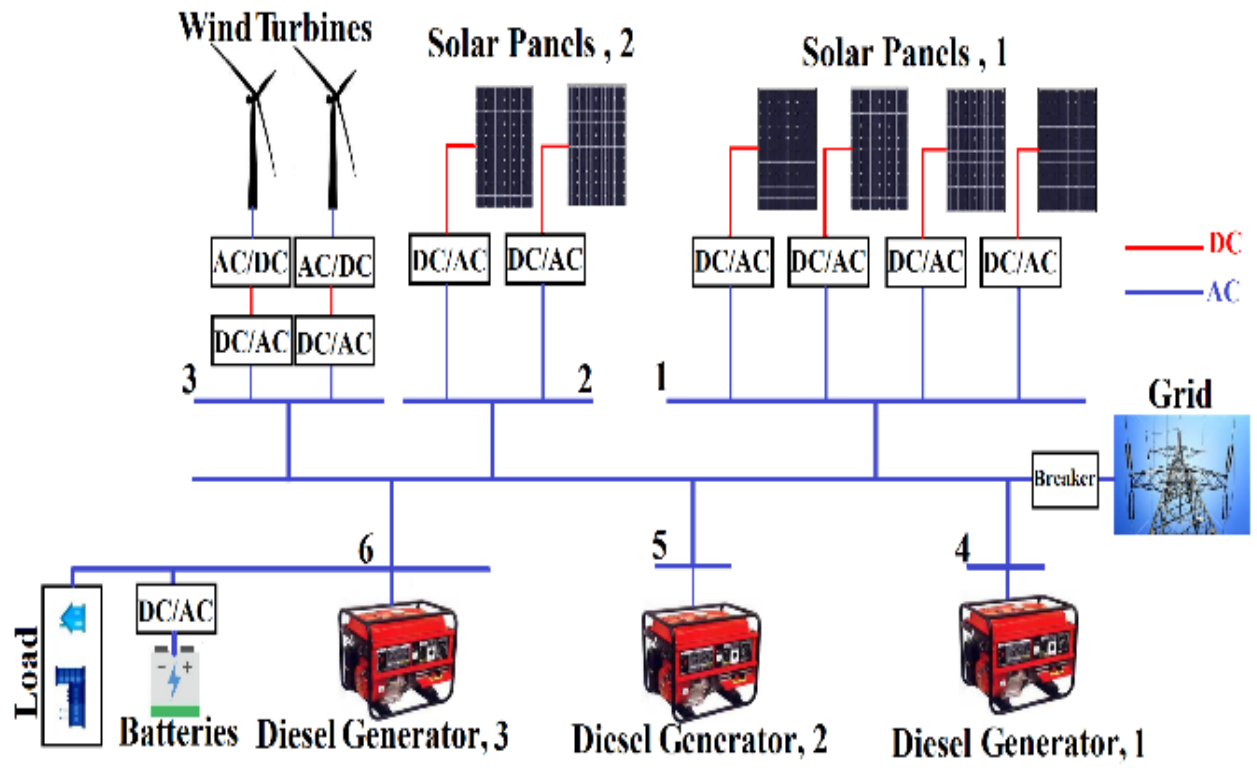

Fig. 1. Schematic of the considered MG.

### B. Modeling of the RESs

Due to the variable generation of the RESs and their probabilistic nature, Weibull and Beta functions are used for the modeling of WTs and PVs power generation, respectively [16]-[17]. First, the average and variance of input data are calculated by using mean and variance in the MATLAB environment [18]; then, the generation of the RESs are produced, using Weibull and Beta distribution functions for five different days. Notably, the output of the distribution functions for each RES in a specific hour is the maximum predicted generation capacity of that source in that hour. The formulas of Weibull and Beta functions are as follows [19]-[20]:

$$g(y') = \frac{1}{B(\alpha,\beta)} \frac{(y'-a)^{\alpha-1}(b-y')^{\beta-1}}{(b-a)^{\alpha+\beta-1}} \tag{1}$$

$$B(\alpha,\beta) = \int_0^1 y'^{\alpha-1}.(1-y')^{\beta-1} dy' \tag{2}$$

$$f(v) = \frac{k_1}{c_1}\left(\frac{v}{c_1}\right)^{k_1-1} \exp\{-\left(\frac{v}{c_1}\right)^{k_1}\} \tag{3}$$

Equations (1) and (2) are related to the Beta distribution function and (3) is related to the Weibull distribution function. Also, $k_1$ and $c_1$ are the Weibull's parameters as well as $\alpha$ and $\beta$ are Beta's parameters. Moreover, $y'$ and $v$ are the input of the functions. Also, $a$ and $b$ are lower and upper limits, respectively.

### C. Load and Price Modeling

Load's accurate forecasting is a difficult job because the behavior of the consumers is stochastic. On the other hand, the energy price in the energy market varies according to the consumer's demand. Hence, the normal distribution function is used to model the load and the energy price uncertainties. Thus, the average and variance of the consumer's demanded load [18] and the energy price [18] are calculated. Then, by using normal distribution function, the hourly load data and the price of electricity is produced for five different days. The normal distribution is defined as follows [21]:

$$h(q,\mu,\sigma^2) = \frac{1}{\sqrt{2\pi\sigma^2}} \exp\left(\frac{-(q-\mu)^2}{2\sigma^2}\right) \tag{4}$$

## III. Problem Formulation

This section has two main parts. The problem constraints are discussed in the first subsection and the objective functions are expressed in the second one.

### A. Constraints

As it is shown in Fig. 1, the intended system consists of several types of equipment, which each of them has its own special constraints. These constraints are classified as follows:

*-Transmission line constraints:* The main grid is connected to the MG through a transmission line. It is obvious that each transmission line is not able to transmit more than a certain amount of power. Therefore, in this work, a maximum value is considered for transmission line capacity. The maximum capacity of the transmission line (MCTL) is selected with respect to the base value of the power ( $S_{base}$ ).

$$P_{trans,t,s} \le MCTL \tag{5}$$

*-DGRs constraints:* DGRs are used to generate power in the MG. For optimal use of DGRs, UC is implemented as well as shutdown and startup costs are considered. Mathematical expressions of these concepts are as follows [22]:

$$y_{g,t,s} - ss_{g,t,s} = v'_{g,t,s} - v'_{g,t-1,s} \tag{6}$$

$$start\ up^{cost}_{g,t} \ge start\ up_g \left[ v'_{g,t,s} - v'_{g,t-1,s} \right]$$
$$start\ up^{cost}_{g,t} \ge 0\ for\ g=1,2,3 \text{ and } t=1,...,24 \tag{7}$$

$$shut^{cost}_{g,t} \ge shut_g \left[ v'_{g,t-1,s} - v'_{g,t,s} \right]$$
$$shut^{cost}_{g,t} \ge 0\ , \text{ for g=1,2,3 and t=1,...,24} \tag{8}$$

The maximum increase rate and minimum decrease rate of the DGRs' power are named up rate and down rate, respectively. The following relations show this statement [22].

$$P_{g,t,s} - P_{g,t-1,s} \le up\ rate_g \tag{9}$$

$$P_{g,t-1,s} - P_{g,t,s} \le down\ rate_g \tag{10}$$

Also, according to UC, the minimum and maximum of power generation of the DGRs are expressed as follows [22]:

$$P_{g,t,s} \le P_g^{\max} \times v'_{g,t,s} \tag{11}$$

$$P_{g,t,s} \ge P_g^{\min} \times v'_{g,t,s} \tag{12}$$

*-BESS constraints:* BESSs have a minimum and maximum state of charge ( $SOC$ ) as well as minimum and maximum rate of charge or discharge; accordingly, the constraints for the $SOC$ and the output power of the BESS are as follows [20]:

$$SOC_{\min} \le SOC_{t,s} \le SOC_{\max} \tag{13}$$

$$0 \le P^{ch,bat}_{t,s} \le P^{ch,bat}_{\max}\ b^{bat}_{t,s} \tag{14}$$

$$0 \le P^{disch,bat}_{t,s} \le P^{disch,bat}_{\max}\ \left(1 - b^{bat}_{t,s}\right) \tag{15}$$

Furthermore, the mathematical expressions of the other constraints of the BESS are presented in (16) and (17). (16) determines how the current $SOC$ will be calculated with respect to the previous hour's $SOC$ . Also, (17) guarantees that the overall $SOC$ level of the BESS does not decrease at the end of a day compared to the previous day [12].

$$SOC_{t,s} = SOC_{t-1,s} + \eta^{ch} \frac{P^{ch,bat}_{t,s}}{S_{base}} - \frac{P^{disch,bat}_{t,s}}{\eta^{disch} S_{base}} \tag{16}$$

$$\sum_{t=1}^{T=24}(SOC_{t,s}-SOC_{t-1,s})\geq 0 \quad (17)$$

*-Constraints of the power exchange:* The maximum power which can be traded between the main grid and the MG is set equal to the maximum capacity of the transmission line (MCTL). Therefore:

$$P_{t,s}^{buy}\leq MCTL\times X_{t,s}^{grid} \quad (18)$$

$$P_{t,s}^{sell}\leq MCTL\times(1-X_{t,s}^{grid}) \quad (19)$$

*-Constraints of the injected power to each bus:* The total generation of the sources connected to each of the buses follows the under-mentioned constraints.

$$Pb_{w,t,s}=\sum_{i=1}^{4}PV_{i,t,s}\ , \qquad (w=1) \quad (20)$$

$$Pb_{w,t,s}=\sum_{i=5}^{6}PV_{i,t,s}\ , \qquad (w=2) \quad (21)$$

$$Pb_{w,t,s}=\sum_{j=1}^{2}WT_{j,t,s}\ , \qquad (w=3) \quad (22)$$

$$Pb_{w,t,s}=P_{g,t,s}\ , \quad (w=4, g=1) \quad (23)$$

$$Pb_{w,t,s}=P_{g,t,s}\ , \quad (w=5, g=2) \quad (24)$$

$$Pb_{w,t,s}=P_{g,t,s}-P_{t,s}^{ch,bat}+P_{t,s}^{disch,bat}-PL_{t,s} \quad (w=6,\ g=3) \quad (25)$$

Equations (20) and (21) indicate that four of the PVs are installed on the first bus and the two of the PVs are installed on the second bus, respectively. Also, (22) shows that there are two installed WTs on the third bus. Furthermore, (23), (24), and (25) demonstrate the DGRs-related buses.

*-Constraints of power balance without DRP:* The power balance equation without considering DRP is stated in (26).

$$\sum_{g=1}^{3}P_{g,t,s}+P_{t,s}^{buy}-P_{t,s}^{sell}+\sum_{i=1}^{6}PV_{i,t,s}+\sum_{j=1}^{2}WT_{j,t,s}-P_{t,s}^{ch,bat}+P_{t,s}^{disch,bat}=PL_{p,t,s} \quad (26)$$

*-Constraints of the RESs:* WTs and PVs have a minimum and maximum generation capacity which are as follows [23]:

$$0\leq PV_{i,t,s}\leq PV_{i,t,s}^{\max} \qquad , \quad i\in\{1,...,6\} \quad (27)$$

$$0\leq WT_{j,t,s}\leq WT_{j,t,s}^{\max} \qquad , \quad j\in\{1,2\} \quad (28)$$

*-Risk assessment constraints:* The relation between RIP and MG's profit is introduced in this section. The MG operators tend to have more profit than a determined lower limit. The $target_p$ is the desired lower limit for the profit of the MG. When the MG's profit is more than $target_p$, it makes the operator satisfied. Otherwise, it is considered a downside risk. Thus, the downside risk constraints for the profit are as follows [13]:

$$\text{if } \ profit_s<\text{target}_{p,s}\ , \ risk_{p,s}=\text{target}_{p,s}-profit_s. \qquad otherwise \ , \ risk_{p,s}=0. \quad (29)$$

The (29) could be expressed as (30) [13].

$$0\leq risk_{p,s}-(target_{p,s}-profit_s)\leq M_p\ .\ (1-W_{p,s}) \qquad 0\leq risk_{p,s}\leq M_p\cdot W_{p,s} \quad (30)$$

$W_{p,s}$ is set equal to 1 when $profit_s<target_{p,s}$. According to the above description, the expected downside risk (EDR) for the profit objective function is as follows [24]:

$$\sum_{s=1}^{5}prob_{p,s}\times risk_{p,s}\leq\lambda_p\ (w_{r,p}-target_p)=\overline{EDR}_p \quad (31)$$

In (31). $prob_{p,s}$ value is set equal to 0.2.

*-DRP constraints:* Nowadays, consumers' role in controlling and management of the power system is increased compared to the past decades. When the MG is connected to the main grid, the MG's operator tries to purchase its required power from the main grid which improves its reliability and economic situation. Also, the MG's operator should implement DRP to obtain the satisfaction of the consumers and have a better economic condition. Hence, in this study, the DRP is performed to investigate the effect of the DRP on the RIP.

By implementing the DRP, the different periods affect each other. Thus, to validate the implemented DRP, in addition to self-elasticity, the cross elasticity is also considered [25]. The main relation of the used DRP is given as follows:

$$PL_{P,t,s}^{LD}=PL_{P,t,s}\times\left\{\begin{array}{l}1+E(t,t).\dfrac{[\rho(t)-\rho_0(t)+A(t)+pen(t)]}{\rho_0(t)}\\ +\sum\limits_{\substack{k=1\\k\neq t}}^{T}E(t,k).\dfrac{[\rho(k)-\rho_0(k)+A(k)+pen(k)]}{\rho_0(k)}\end{array}\right\} \quad (32)$$

When customers consume more than the contract, they will be penalized. The amount of penalty is assumed to be zero.

The relations of the limitation of the customers' participated power as well as the new demanded load of the consumers after implementation of the DRP are as follows [23]:

$$-0.25\,PL_{p,t,s}\leq PL_{p,t,s}^{D}\leq 0.25\,PL_{p,t,s} \quad (33)$$

$$PL_{p,t,s}^{LD}=PL_{p,t,s}+PL_{p,t,s}^{D} \quad (34)$$

The consumers do not eliminate the load; they just transfer it from one period to another. So, it can be written as follows [23]:

$$\sum_{t=1}^{T}PL_{p,t,s}^{D}=0 \quad (35)$$

Also, the DRP cost function is given as follows [23]:

$$Cost_{DRP}=\sum_{t=1}^{T}\left(PL'^{\,D}_{p,t,s}\,A_t\right)\ ,\ PL'^{\,D}_{p,t,s}=|\,PL_{p,t,s}^{D}\,| \quad (36)$$

*-Constraints of power balance with DRP:* Due to the execution of DRP in the considered MG, the power balance equation in this state differs from the previous one. Thus, the power balance equation to investigate the DRP effect on the RIP is as follows:

$$\sum_{g=1}^{3}P_{g,t,s}+P_{t,s}^{buy}-P_{t,s}^{sell}+\sum_{i=1}^{6}PV_{i,t,s}+\sum_{j=1}^{2}WT_{j,t,s}-P_{t,s}^{ch,bat}+P_{t,s}^{disch,bat}=PL_{p,t,s}^{LD} \quad (37)$$

## B. Objective Functions

Assessing the effect of the DRC on the profit and surveying the impact of the DRP on the RIP in MGs are considered in this work. Comparing the results with considering DRC (CDRC) and without considering DRC (WCDRC) will determine the effect of the downside risk constraints and DRP on the defined parameters. Due to the existence of various goals in this study, objective functions are divided into the following subsections.

*-Objective function without DRP:* This objective function is used to investigate the MG's profit in the CDRC and WCDRC cases which is indicated in (38) and (39).

$$ZZ_s^{profit}=\sum_{t=1}^{24}P_{t,s}^{sell}\times C_{t,s}^{sell}-\sum_{t=1}^{24}P_{t,s}^{buy}\times C_{t,s}^{buy}-\sum_{t=1}^{24}\sum_{g=1}^{3}(b_g\times P_{g,t,s}+V'_{g,t,s}\times c_g) -\sum_{t=1}^{24}\sum_{g=1}^{3}shut_g\times SS_{g,t,s}-\sum_{t=1}^{24}\sum_{g=1}^{3}start\ up_g\times y_{g,t,s} \quad (38)$$

$$\overline{Z}_{profit}=\sum_{s=1}^{5}prob_s\times ZZ_s^{profit} \quad (39)$$

*-Objective function with DRP:* This objective function is utilized to investigate the effect of the DRP in this proposed model in the WCDRC and CDRC cases.

$$ZZ_s^{profit,DRP} = \sum_{t=1}^{24} P_{t,s}^{sell} \times C_{t,s}^{sell} - \sum_{t=1}^{24} P_{t,s}^{buy} \times C_{t,s}^{buy} - \sum_{t=1}^{24}\sum_{g=1}^{3}(b_g \times P_{g,t,s} + V'_{g,t,s} \times c_g) - \sum_{t=1}^{24}\sum_{g=1}^{3} shut_g \times SS_{g,t,s} - \sum_{t=1}^{24}\sum_{g=1}^{3} start\ up_g \times y_{g,t,s} - \sum_{t=1}^{24}\left(PL'^{D}_{p,t,s} \times A_t\right) \quad (40)$$

$$\bar{Z}_{profit}^{DRP} = \sum_{s=1}^{5} prob_s \times ZZ_s^{profit,DRP} \quad (41)$$

## IV. Proposed Optimization Method

The presented optimization model for this study is a MIP problem. The GAMS by using CPLEX solver and MATLAB software are utilized to solve this problem. The optimization method is briefly described in the following steps:

*Step 1:* Importing the initial data, which are not dependent on the scenarios and the considered objective functions.

*Step 2:* Stochastic production of data such as MG's load, purchasing and selling price of the electricity, generated power of RESs are performed in MATLAB software and the obtained data used in the GAMS software. The study is implemented for 5 days to make a proper assessment of the proposed model. Thus, five groups of data are produced in MATLAB software and sent to GAMS software as initial data. Each of the above-mentioned groups is called a scenario. The purpose of creating several scenarios is performing stochastic scheduling.

*Step 3:* Study on the MG to investigate the MG's profit in the WCDRC case. In this section, the 1st-7th constraints groups and then the first objective function, considering (38) and (39), are used. Note that the objective function, initial data, and constraints are considered for each of the five scenarios. Then, the MG's profit is calculated.

*Step 4:* Study on the MG to investigate the MG's profit in the CDRC case. The 1st-8th constraints groups and the first objective function, considering (38) and (39), are utilized in this step. Like the third step, the studies are conducted for 5 scenarios and the profit of MG is calculated for the CDRC case. The results of this step are compared with the results of step 3.

*Step 5:* Study on the MG for analyzing the effect of the DRP on the MGs' profit values in the WCDRC case. The 1st-5th, 7th, 9th, and 10th constraints groups with the second objective function, considering (40) and (41), are utilized in this step. The second objective function, considering (40) and (41), with the introduced constraints are studied and the MG's profit for each scenario is calculated.

*Step 6:* Study on the MG for analyzing the effect of the DRP on the MGs' risk and profit values at the same time in the CDRC case. The 1st-5th, and 7th-10th constraints groups with the second objective function, considering (40) and (41), are used in this step. The amount of risk and profit is calculated for each of the scenarios in this step. The results of this step are compared with the results of step 5. After the completion of this step, the results of steps 3-6 are compared.

*Step 7:* Study on the MG for analyzing the robustness and efficiency of the proposed model with DRP consideration by performing the sensitivity analysis in the WCDRC case and CDRC case. The 1st-10th constraints groups and two objective functions are used. Finally, the results are saved to get analyzed with the results of the other steps.

## V. Results and Discussion

In this paper, two main objectives with several subsections are investigated. The first objective is profit maximization as well as risk minimization and the second objective function is surveying the influence of the DRP on the profit and risk of the MGs. Minimizing RIP is the secondary aim of all studies. To achieve these objectives, the steps which are introduced in the optimization steps section are followed. Thus, the first step is performed, and the required initial data are produced. The maximum generation capacity of the overall WTs is 42 kW and the maximum generation capacity of the overall PVs connected to the first and second buses are 32 kW and 16 kW, respectively. These generation capacities are selected according to the MG's demanded load and the used reference. The base value of the power ( $S_{base}$ ) and the maximum capacity of the transmission line (MCTL) is set equal to 50 kW and 37.5 kW (0.75 per-unit (p.u.), respectively. The initial $SOC$ of the BESSs is assumed 70%. The values of $SOC_{\min}$ , $SOC_{\max}$ , $P_{\min}^{bat}$ , and $P_{\max}^{bat}$ are assumed 20%, 100%, -50 kW and 50 kW, respectively. The charge efficiency and discharge efficiency of the BESS are set equal to 100%. Also, it is mentioned that there is just one DGR in each of the buses. The characteristics of the utilized DGRs are shown in Table I. These data are obtained from modifying the DGRs' coefficient presented in [22].

TABLE I
Characteristics Of The Dgrs

| $g$ | $b_g$ | $P_g^{\min}[kW]$ | $P_g^{\max}[kW]$ | $down\ rate_g[kW]$ | $up\ rate_g[kW]$ |
|---|---|---|---|---|---|
| 1 | 0.7 | 0 | 4 | 3 | 3 |
| 2 | 0.25 | 0 | 6 | 5 | 5 |
| 3 | 0.5 | 0 | 9 | 8 | 8 |

Since the DGRs' capacity is low, the linear cost function is used. Table I includes the coefficient of DGRs' cost function, minimum and maximum power generation of each DGR as well as up rate and down rate of them. $c_g$ is the coefficient of DGRs' cost function, and it is set equal to zero. Also, UC is performed on DGRs. It should be noted that all the studies are performed in 5 random days of a year. All the studies in this article have been done for a short period, which does not include costs including initial investments and periodic repairs. For this reason, fuel costs of the DGRs are considered.

When the first step is accomplished, step two is performed. There are some parameters with stochastic values such as the generated power of the RESs, MG's load, and energy price; in which the Beta, Weibull, and Normal probabilistic distributions are used to create them [18]. Also, the MG's considered load in 5 random days of a year is presented in Fig. 2.

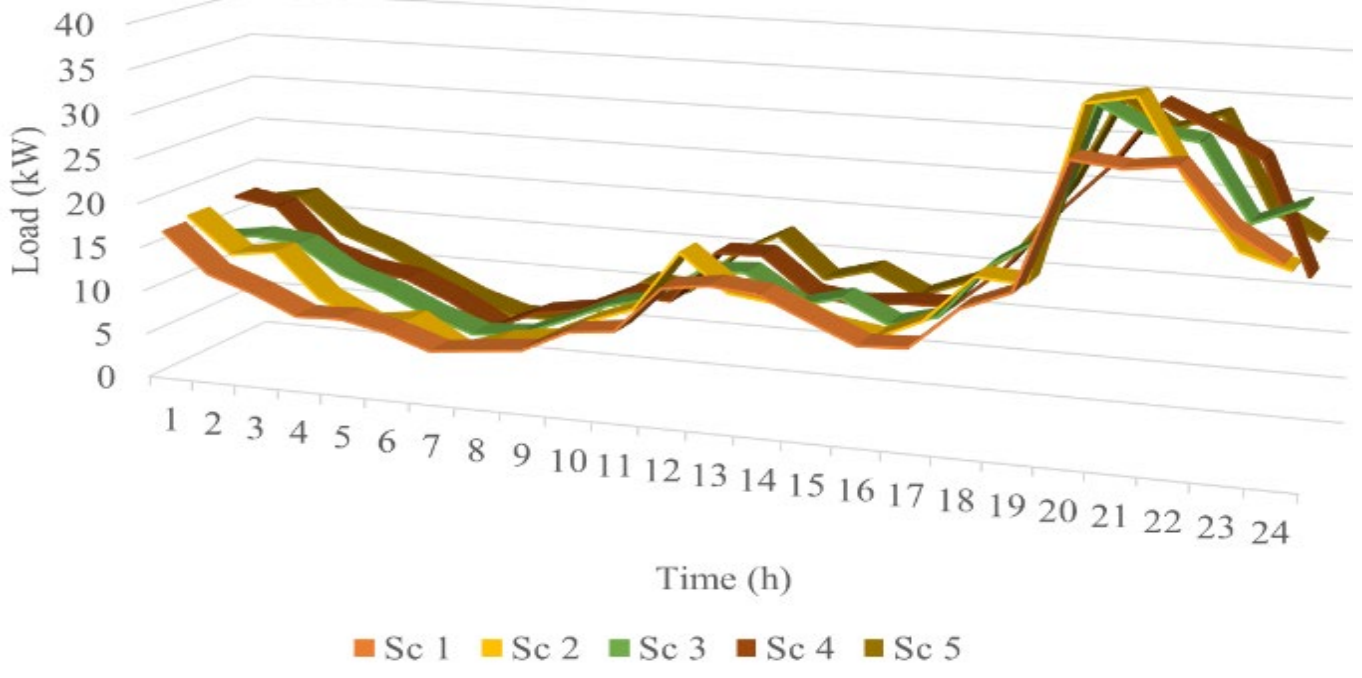

Fig. 2. MG's load.

According to the mentioned steps in the optimization section, steps one and two are accomplished and the initial data are produced. In the next step, the simulations are performed to profit maximization in the WCDRC case. The amount of the profit, RIP for each scenario as well as the average profit and RIP are shown in Table II. The average profit for the mentioned 5 scenarios is \$72.6946 and the average RIP is \$0.5979. The target value for the fourth and fifth steps is set equal to \$72.6946 based on the average profit. In the fourth step, the studies are performed to maximize the MG's profit in the CDRC case. The results are shown in Table III and Table IV.

The profits for each scenario with different $\lambda_p$ are illustrated in Table III. Also, Table IV indicates the average profit, total RIP, average RIP, average profit's reduction, and average RIP's reduction in comparison to the WCDRC case for each $\lambda_p$. Comparison of Tables II, III, and IV show that the risk is reduced significantly versus a very low reduction of the profit percentage. For example, Table IV shows that for $\lambda_p = 0.7$ the RIP is 29.9987% reduced in comparison to the WCDRC; while, the reduction of the profit is only 0.2462%. Moreover, increasing the value of $\lambda_p$ leads to an increase in the average profit and RIP reduction, but the reduction rate of RIP is more than the rate of the average profit reduction. For instance, considering $\lambda_p = 0.99$ as well as $\lambda_p = 0.7$ in Table IV, it is observed that the average RIP is reduced by 29.0011% against 0.2445% reduction of the average profit, which can be neglected. The comparison schematics of the average RIP and average profit for different $\lambda_p$ s as well as CDRC case versus WCDRC case are shown in Figs. 3 and 4.

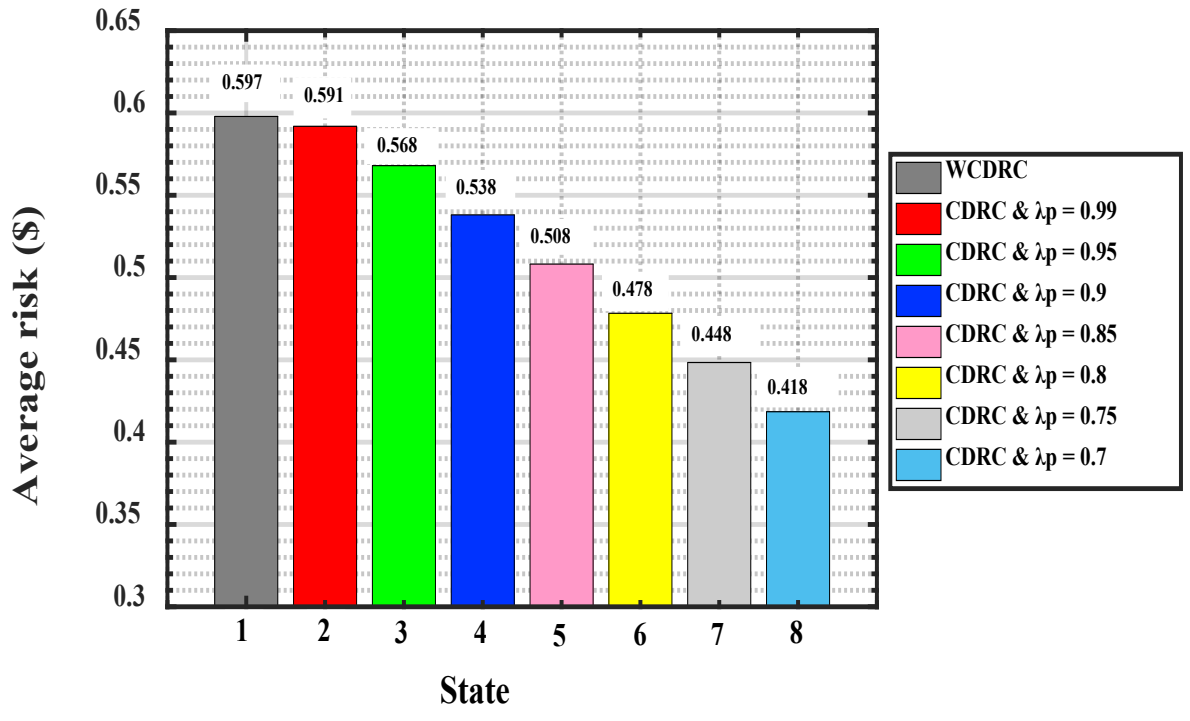

Fig. 3. Comparison schematic between RIP with various $\lambda_p$ s.

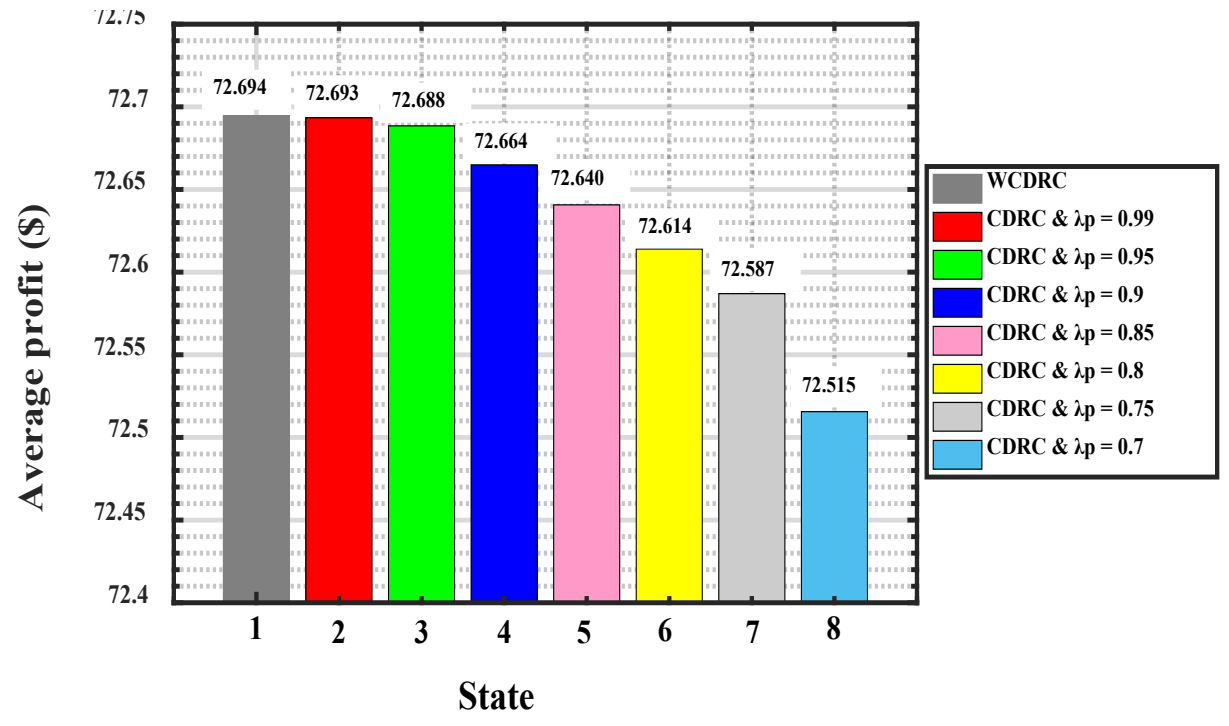

Fig. 4. Comparison schematic between average profit with various $\lambda_p$ s.

TABLE II
RESULTS OF THE PROFIT MAXIMIZATION IN THE WCDRC CASE [\$]

| Scenario | Profit | RIP |
|---|---|---|
| 1 | 72.10782 | 0.5868 |
| 2 | 73.42139 | 0 |
| 3 | 70.29204 | 2.4026 |
| 4 | 74.01032 | 0 |
| 5 | 73.64145 | 0 |
| Average | 72.69460 | 0.5979 |

TABLE III
EACH OF THE SCENARIO'S PROFIT WITH DIFFERENT $\lambda_p$ S FOR CDRC CASE [\$]

| $\lambda_p$ | Sc 1 | Sc 2 | Sc 3 | Sc 4 | Sc 5 |
|---|---|---|---|---|---|
| 0.99 | 72.1078 | 73.3854 | 70.3219 | 74.0103 | 73.6414 |
| 0.95 | 72.1078 | 73.2413 | 70.4414 | 74.0103 | 73.6414 |
| 0.9 | 72.1078 | 72.9739 | 70.5910 | 74.0103 | 73.6414 |
| 0.85 | 72.1078 | 72.7035 | 70.7404 | 74.0103 | 73.6414 |
| 0.8 | 72.1078 | 72.6947 | 70.8899 | 73.7362 | 73.6414 |
| 0.75 | 72.1085 | 72.6947 | 71.0386 | 73.4518 | 73.6414 |
| 0.7 | 72.2581 | 72.6947 | 71.0386 | 72.9453 | 73.6414 |

TABLE IV
RESULTS OF THE MG'S PROFIT MAXIMIZATION IN THE CDRC CASE AS WELL AS COMPARISON WITH WCDRC CASE

| $\lambda_p$ | Average profit [\$] | Total RIP [\$] | Average RIP [\$] | Average profit reduction [%] | Average RIP reduction [%] |
|---|---|---|---|---|---|
| 0.99 | 72.6934 | 2.9596 | 0.5919 | 0.0017 | 0.9976 |
| 0.95 | 72.6885 | 2.8401 | 0.5680 | 0.0085 | 4.9940 |
| 0.9 | 72.6649 | 2.6906 | 0.5381 | 0.0409 | 9.9969 |
| 0.85 | 72.6407 | 2.5411 | 0.5082 | 0.0741 | 14.9949 |
| 0.8 | 72.6140 | 2.3916 | 0.4783 | 0.1109 | 19.9978 |
| 0.75 | 72.5870 | 2.2422 | 0.4484 | 0.1480 | 24.9958 |
| 0.7 | 72.5156 | 2.0926 | 0.4185 | 0.2462 | 29.9987 |

The effect of DRP on profit and RIP has been investigated in steps 5 and 6, respectively. In the fifth step, the goal is to maximize the profit of the MG by considering DRP and without considering the risk. The results of this section are shown in Table V. In steps 5 and 6, the maximum rate of consumer participation in DRP is assumed to be 25%. Table V shows the profit and risk values in each scenario and the average profit and RIP. Also, the target value for steps 5 and 6 is \$75.1807.

TABLE V
SURVEY RESULTS OF THE MG'S PROFIT CONSIDERING DRP AND WCDRC [\$]

| Scenario | Profit | RIP |
|---|---|---|
| 1 | 74.1715 | 1.0093 |
| 2 | 77.4795 | 0 |
| 3 | 72.5267 | 2.6541 |
| 4 | 75.7385 | 0 |
| 5 | 75.9878 | 0 |
| Average | 75.1807 | 0.7326 |

To make a detailed assessment, step 6 is performed, and the results are compared with step 5 and Table V. In the sixth step, the goal is to examine the MG's profit with DRP and risk constraints. The results are shown in Tables VI and VII. Table VI indicates the profit of the MG for different scenarios and different $\lambda_p$ s. Also, for achieving the better and more accurate comparisons, the average profit, total RIP, average RIP, percentage of RIP reduction, and the percentage of profit reduction for various $\lambda_p$ are shown in Table VII. Comparison of the results of Tables V, VI, and VII show that by decreasing $\lambda_p$, RIP has been significantly reduced, and profit value has decreased. Also, according to Table VII the rate of RIP reduction is much faster than the rate of reduction in profit. For example, if two states $\lambda_p = 0.99$ and $\lambda_p = 0.7$ are considered. In Table VII, it is concluded that against a 29.0485% reduction in RIP, the profit value has been faced with only a 0.0493% decrease. The schematic representation of the comparison between the average profit and the average RIP for the sixth step is shown in Figs. 5 and 6, respectively. As shown in Fig. 5 and Fig. 6, by decreasing $\lambda_p$, the slope of risk reduction is faster than the rate of decline in profit. According to Figs. 5 and 6, for $\lambda_p$ higher than 0.8, no significant changes have been made, but the RIP has been steadily decreasing, which could be attractive for the MGs operators. The goal of this work is the maximization of the MG's profit and minimization of the RIP. There is an intrinsic direct relationship between RIP and profit. For this reason, it is not possible to reach the ideal point, but it is possible to find points in which the changes in the profit reduction are very low and the RIP reduction is so high, and these points are important for the MG's operator. Similarly, for $\lambda_p$ less than 0.8, the reduction in profit is faster while the slope of RIP changes has maintained a downward trend (for larger $\lambda_p$ s). Therefore, the most reasonable result that can be achieved is that the MG's operator checks points above 0.8. On the other hand, sometimes the MG's operator decides to minimize the RIP and operate its MG with a $\lambda_p$ below 0.8 point in which it should pay the cost of reducing the MG's profit. Meanwhile, this reduction in profit is not so high, therefore the MG's operator is always interested in reducing the risk in the MGs.

TABLE VI

SCENARIOS PROFIT WITH DIFFERENT $\lambda_p$ S IN THE CDRC CASE WITH DRP [$]

| $\lambda_p$ | Sc 1 | Sc 2 | Sc 3 | Sc 4 | Sc 5 |
|---|---|---|---|---|---|
| 0.99 | 74.1715 | 76.8600 | 72.5633 | 76.3213 | 75.9878 |
| 0.95 | 74.1715 | 76.8600 | 72.7100 | 76.1747 | 75.9878 |
| 0.9 | 74.1715 | 76.8600 | 72.8933 | 75.9913 | 75.9878 |
| 0.85 | 74.1715 | 76.8600 | 73.0768 | 75.8079 | 75.9878 |
| 0.8 | 74.1715 | 76.7696 | 73.2602 | 75.7097 | 75.9878 |
| 0.75 | 74.1715 | 76.6167 | 73.4436 | 75.5907 | 75.9878 |
| 0.7 | 74.1715 | 76.3416 | 73.6269 | 75.5907 | 75.9878 |

TABLE VII

MG'S PROFIT IN CDRC CASE AND COMPARISON OF WCDRC CASE WITH DRP

| $\lambda_p$ | Average profit [$] | Total RIP [$] | Average RIP [$] | Average profit reduction [%] | Average RIP reduction [%] |
|---|---|---|---|---|---|
| 0.99 | 75.1808 | 3.6272 | 0.7254 | 0 | 0.9868 |
| 0.95 | 75.1808 | 3.4800 | 0.6960 | 0 | 5.0049 |
| 0.9 | 75.1808 | 3.2967 | 0.6593 | 0 | 10.0093 |
| 0.85 | 75.1808 | 3.1132 | 0.6226 | 0 | 15.0178 |
| 0.8 | 75.1797 | 2.9299 | 0.5859 | 0.0014 | 20.0223 |
| 0.75 | 75.1620 | 2.7464 | 0.5493 | 0.0249 | 25.0308 |
| 0.7 | 75.1437 | 2.5631 | 0.5126 | 0.0493 | 30.0353 |

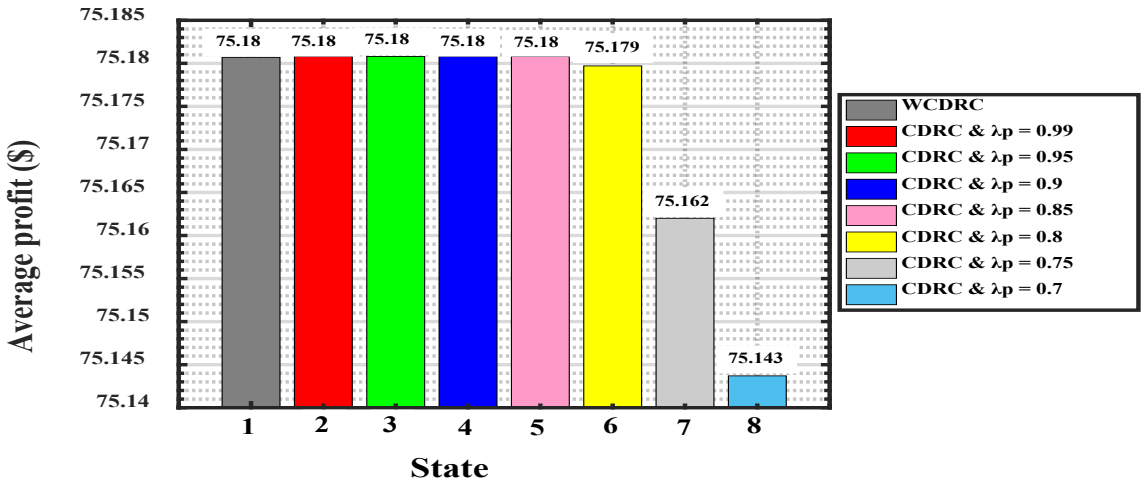

Fig. 5. Average profit with and without risk constraint and with DRP.

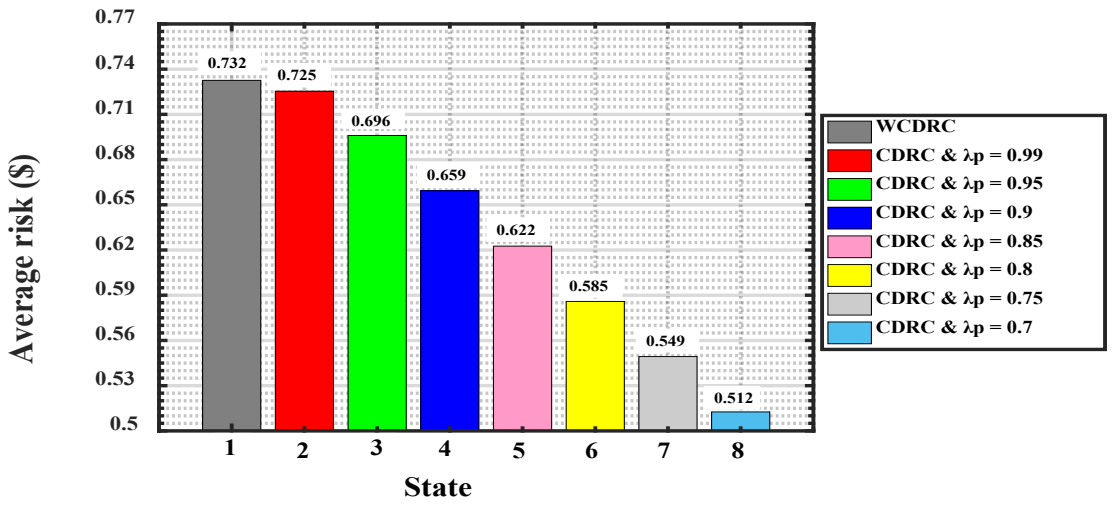

Fig. 6. RIP with and without considering RIP state with DRP.

The schematic comparison of the results of steps 3 and 5 is shown in Figs. 7 and 8. These figures show changes in average profit and average RIP of implementing and not implementing the DRP without considering risk, respectively.

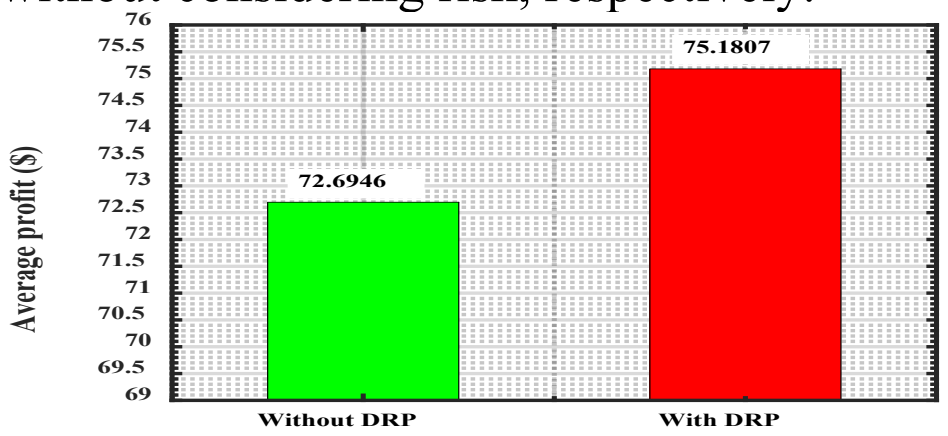

Fig. 7. Average profit with and without DRP and WCDRC.

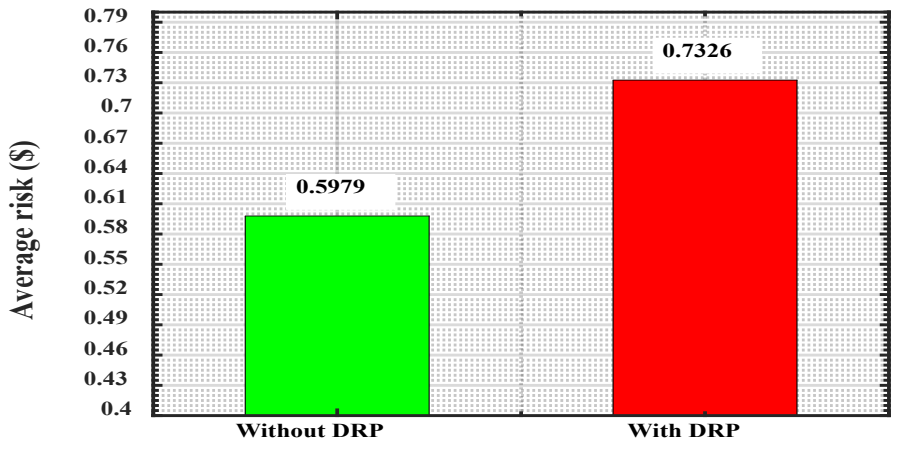

Fig. 8. RIP with and without considering DRP and WCDRC.

As shown in Figs. 7 and 8, by applying DRP without risk, the average profit of the MG and the average risk rate increase compared to the without DRP and without the risk. Also, according to Figs. 7 and 8, by applying the DRP, the percentage of increase in RIP is higher than the increasing percentage in profit. The increase in RIP was about 22.5288% and the profit was 3.4199% according to Figs. 7 and 8, respectively. This means that the sensitivity of the RIP is greater than the MG's profit, due to the implementation of DRP. The schematic of the compared results of steps 4 and 6 is also shown in Figs. 9 and 10. These figures show, respectively, average changes in profit and average RIP changes for performing and not performing the DRP by considering the risk for different $\lambda_p$ s. According to Figs. 9 and 10, similar to WCDRC scenarios, when applying DRP by considering risk, the average profit of the MG and the average RIP increase as compared to the without DRPs and CDRC. According to Fig. 10, for high values of $\lambda_p$ s, the average increase in the RIP is higher than $\lambda_p$ s with low values while these sensible changes are not seen in the profit.

According to Figs. 7-10, RIP sensitivity to with and without the implementation of DRP is higher than the profit sensitivity. So that by changing the $\lambda_p$ and rate of loads participation in the DRP, the RIP changes are more sensible than the changes in profit. Also, RIP changes are greater than the profit changes.

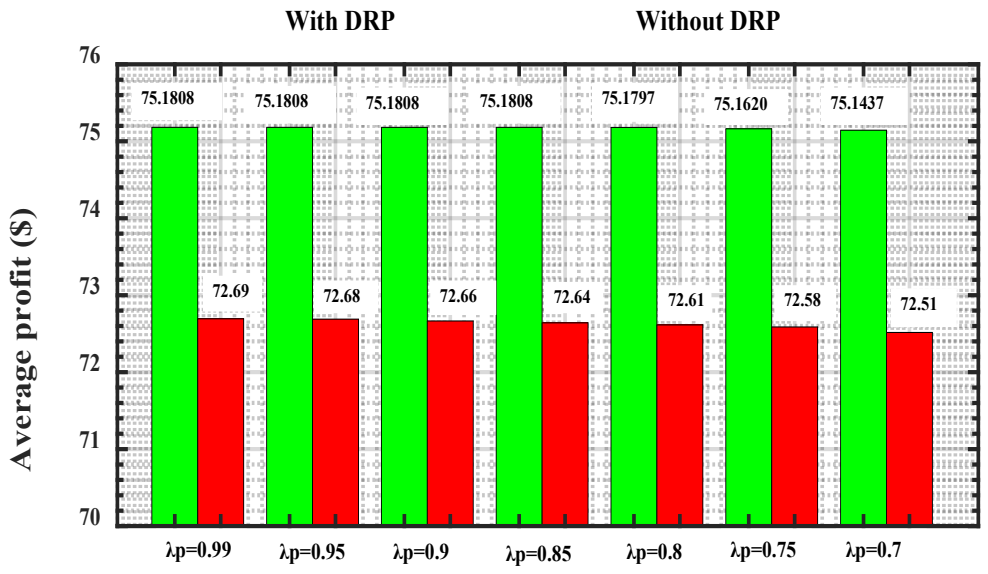

Fig. 9. Average profit's with and without considering DRP states and CDRC.

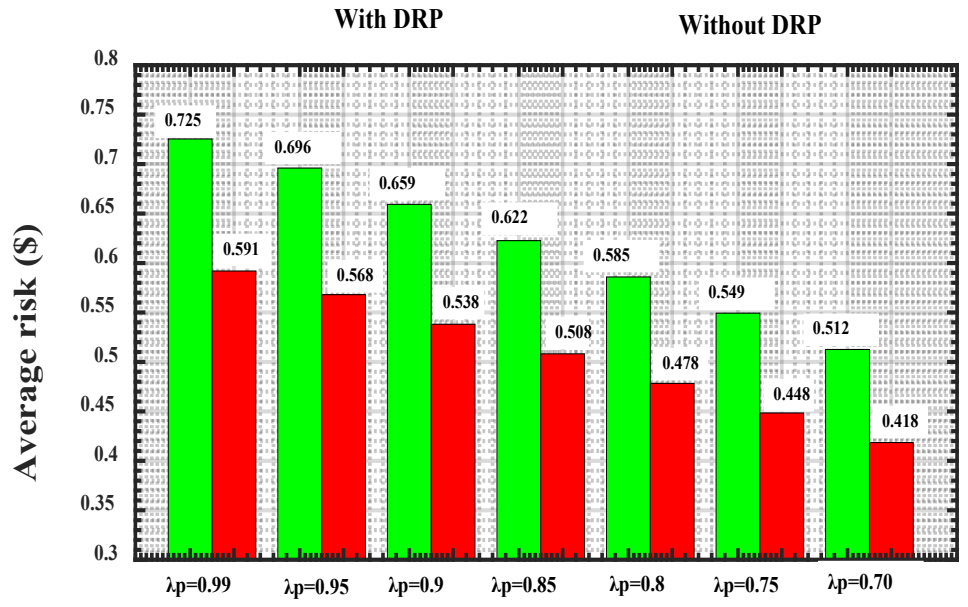

Fig. 10. RIP with and without considering DRP states and CDRC.

To this stage, it was examined how DRP and risk constraints could have an impact on the profit and RIP of the MGs. Also, the rate of participation in the DRP is assumed to be fixed at 25%. At this stage and in the seventh step, the rate of participation of the loads is also changed, and it should be examined what effects it can have on the profit and the RIP of the MG. In fact, it is intended that a sensitivity analysis is carried out. For this purpose, in three completely separate modes, it is assumed that the rate of participation in the DRP is 20%, 25%, and 30%, and for each one, it is examined which results will be in the presence and absence of downside risk constraints. It should be noted that the target value was 73,9981 $, 75.1808 $, and 76.4125 $, respectively, when loads participation in the DRP are 20%, 25%, and 30%, respectively. In step seven it is assumed that the risk constraints in the modeling of the problem are not considered and the rate of participation is changed. The results of this study are schematically shown in Figs. 11 and 12. Figs. 11 and 12 show changes in average RIP and changes in profit according to the rate of the load's participation percentage in DRP, respectively. Based on Figs. 11 and 12, by increasing the participation of loads in DRP, the average profit, and average RIP increase. It is also observed that with the same change in the participation rate (5%), the slope of the average profit changes is regular and almost linearly changed, but average RIP changes are irregular and nonlinear, but their changes have always been ascending. Thus, Figs. 11 and 12 represent that loads participation in DRP should not be high. For example, when the value of the participation varies from 25% to 30%, the profit value is approximately the same as the previous step (between 20% and 25%), but the RIP has increased. These choices are usually selected by the MG's operator to achieve higher profit.

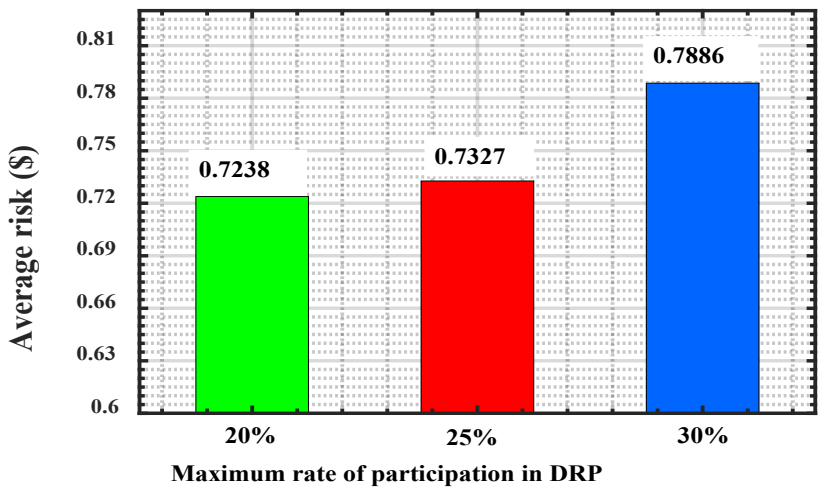

Fig. 11. Average RIP in WCDRC considering loads participation with DRP.

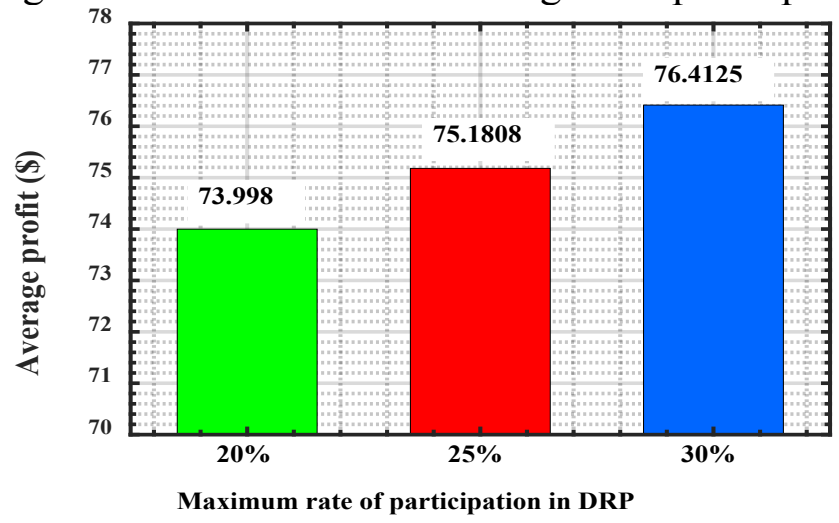

Fig. 12. Average profit in WCDRC considering loads participation with DRP.

In the next step, The risk constraints are considered in the modeling of the problem and the rate of participation is changed to investigating the changes in the MG's profit and RIP. The results of this study are schematically shown in Figs. 13 and 14.

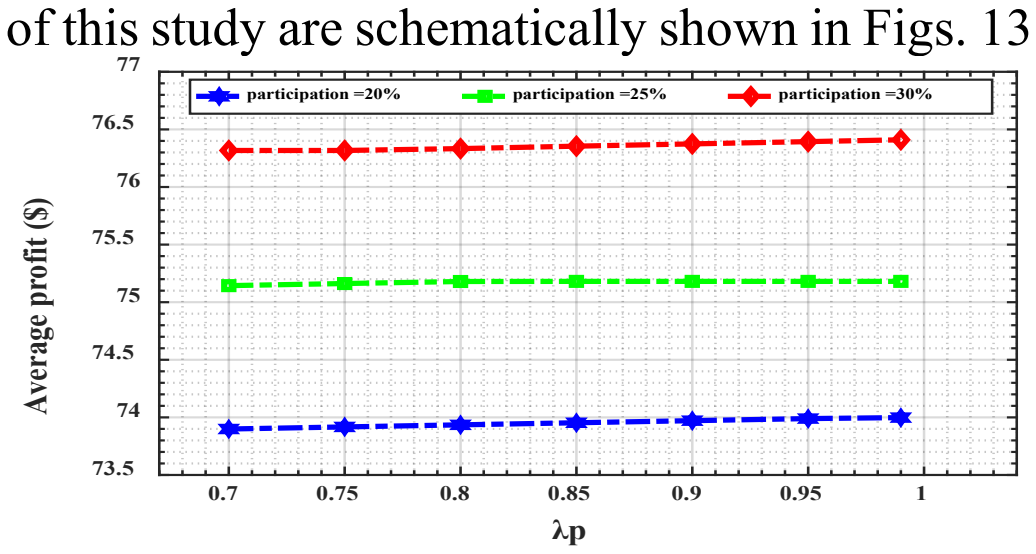

Fig. 13. Average profit in CDRC for loads participation and changing $\lambda_p$.

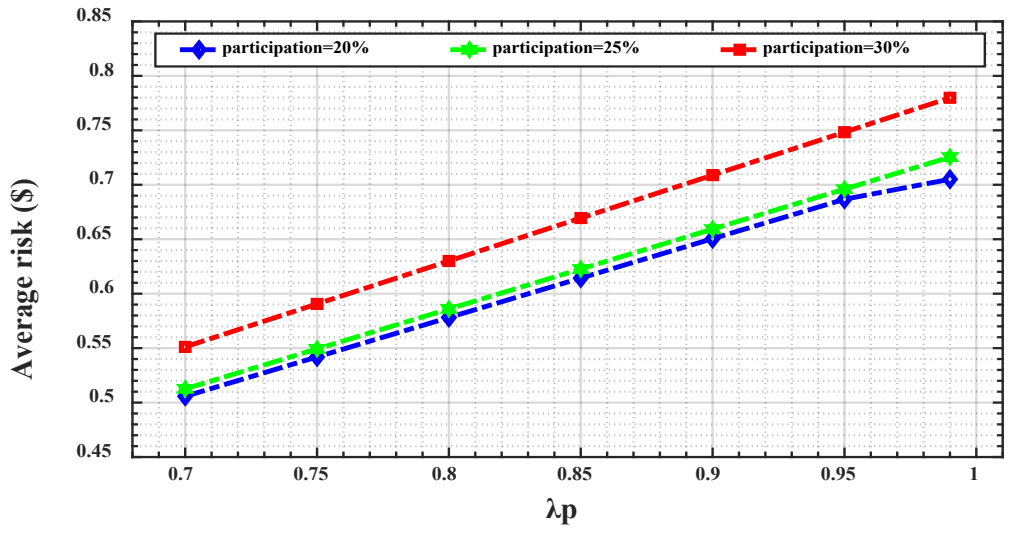

Fig. 14. Average RIP in CDRC for loads participation and changing $\lambda_p$.

These figures show, respectively, changes in average profit and average risk according to the changes in the rate of the load's participation in DRP and the variations in the $\lambda_p$. It should be noted that in Figs. 13 and 14, the word participation is the maximum permitted rate of participation of loads in DRP. Based on Figs. 13 and 14, increasing the participation rate in DRP increases the average profit and average RIP. Also, with the increase of $\lambda_p$, the average profit and average RIP will increase. Also, with the same change in the participation rates, the slope of the average profit changes is slower than the slope of RIP changes, but the trend of these changes has been almost upward. Regarding Figs. 13 and 14, when the value of the load's participation varies from 25% to 30%, the profit changes have roughly regulated. These changes are nearly the same as the previous step (between 20% and 25%). Also, these changes are incremental, but the changes in RIP have increased a lot.

The goal of this study is to examine the parameters in which their changes are effective in profit and RIP changes, and this analysis and sensitivity analysis is carried out in 7 steps. The first and second steps are obtaining the initial data, and the next five steps are the main steps. In the third and fourth steps, it is examined that the presence of or non-presence of risk constraints could have any effect on the profit and RIP when the MGs loads not participating in the DRP. In the fifth and sixth steps, the purpose of the study is to investigate the profit and RIP changes with the participation of loads in DRP and in the presence or non-presence of risk constraints. In the seventh step, sensitivity analysis is also carried out in which how by changing the value of the load's participation rate and the presence and non-presence of risk constraints the profit of the MG and RIP change. In this survey, it is found that generally there is always a direct relationship between RIP and profit. Of course, their relationship is not linear, and the value of their changes is not related to each other, but their manner is similar. For example, both are incremental, but their increase is not the same. Applying the risk constraints always reduces the profit and RIP, and this expression exists for both DRP and without DRP modes. Studies have also shown that by implementing the DRP, profit and RIP are increased. Also, when the participation of loads in DRP is increased, the profit and RIP are increased. Generally, these surveys help the MG's operator and improve the profit and minimize the RIP in different states.

## VI. Conclusion

The optimal power scheduling of the resources in an MG is performed to minimize the risk. The main objectives of this work are profit maximization as well as investigating the effect of the DRP on the risk and the profit of the MGs. The secondary goals are the risk minimization and optimal scheduling of the energy sources. UC is performed on the DGRs. First, the profit of the MGs for both WCDRC and the CDRC is investigated. The results of this MIP problem show that for the CDRC case the profit is slightly reduced, but RIP is reduced significantly; in one of the cases, RIP is reduced by 29.9987%; whereas the average profit is only reduced by 0.2462%. Without considering DRC, the simulation results show that DRP has significant impacts on the risk and the profit of the MGs. Then, the DRCs are added to the problem, and the optimization is performed. The comparison between WCDRC and CDRC cases shows that the average profit slightly decreases while the average RIP significantly decreases. Therefore, CDRC and applying DRP at the same time lead to a dramatic reduction of RIP. After that sensitivity analysis is executed, and the results confirm the proper performance of the proposed model. Thus, some constraints and techniques are suggested which lead to a significant reduction in the risk and a notable increment in the profit of the MGs.